\documentclass[11pt, oneside]{amsart}
\usepackage{amsmath}
\usepackage{amsthm}
\usepackage{amssymb} 
\usepackage[mathscr]{euscript}
\usepackage{amsmath}
\usepackage{amsthm}
\usepackage{amssymb}
\usepackage{fancyhdr}
\usepackage{enumerate}
\usepackage{amscd}
\usepackage[all]{xy}
\usepackage{graphicx}

\usepackage{amsaddr}

\usepackage{xcolor}

\usepackage[normalem]{ulem}

\newtheorem{thm}{Theorem}

\newtheorem{cor}[thm]{Corollary}

\theoremstyle{definition}
\newtheorem{definition}[thm]{Definition}

\theoremstyle{remark}
\newtheorem{remark}[thm]{Remark}

\DeclareMathOperator{\tr}{tr}
\DeclareMathOperator{\Div}{div}
\DeclareMathOperator{\ric}{Ric}
\def\rr{\mathbb{R}}

\renewcommand\S{\Sigma}
\renewcommand\l{\lambda}
\renewcommand\th{\theta}
\newcommand\calH{{\mathcal{H}}}
\newcommand\beq{\begin{equation}}
\newcommand\eeq{\end{equation}}
\newcommand\ben{\begin{enumerate}}
\newcommand\een{\end{enumerate}}
\newcommand\bit{\begin{itemize}}
\newcommand\eit{\end{itemize}}

\newcounter{mnotecount}

\begin{document}

\title[A note on the positive mass theorem with boundary]{A note on the positive mass theorem \\ with boundary}

\author{Gregory J. Galloway}
\address{Department of Mathematics, University of Miami \\  Coral Gables, FL 33124, USA}

\author{Dan A. Lee}
\address{Department of Mathematics, Graduate Center and Queens College, CUNY \\ 365 Fifth Avenue,
New York, NY 10016, USA}
\email{dan.lee@qc.cuny.edu}

\begin{abstract}
In this short note we show how one can use established results to prove various versions of the positive mass theorem for initial data sets with boundary, in dimensions less than 8.
\end{abstract}


\maketitle

The positive mass theorem for asymptotically flat initial data sets without boundary (sometimes called the ``spacetime positive mass theorem'') was established in the spin case by E.~Witten in~\cite{Witten1981} and in dimensions less than 8 in~\cite{EichmairHuangLeeSchoen}, the latter of which was based on~\cite{SchoenYau1979.1, Schoen1989, Schoen2005} and strengthened the positive energy theorem of R.~Schoen and S.-T.~Yau in dimension~3~\cite{SchoenYau1981}. The general case was treated in~\cite{Lohkamp2016}. It is generally believed that the theorem applies to initial data sets \emph{with boundary} as long as one assumes that the boundary is \emph{weakly outer trapped}. This was first observed and verified in the spin case in~\cite{GibbonsHawkingHorowitzPerry, Herzlich1998}. In the time-symmetric case, it follows from~\cite{Miao}. More recently,  S.~Hirsch, D.~Kazaras, and M.~Khuri~\cite{HirschKazarasKhuri} have discovered a new proof of the positive mass theorem for 3-dimensional initial data sets, based on ideas from~\cite{Stern}, and their results imply a positive mass theorem for initial data sets with weakly outer trapped boundary. Our goal here is to explain how the technique of~\cite{EichmairHuangLeeSchoen} can be generalized to the case of outer trapped boundary, and how that fact can then be used to apply to other situations.

We begin by reviewing some of the relevant definitions. Given initial data $(g, k)$ on a manifold $M$, we define
\begin{align*}
\mu&:= \tfrac{1}{2}\left[ R_g + (\tr_g k)^2 -|k|_g^2\right],\\
J^i&:= (\Div_g k)^i - \nabla^i (\tr_g k),
\end{align*}
and say that the \emph{dominant energy condition (or DEC)} holds at a point $p\in M$ if $\mu\ge |J|_g$ at $p$. 

Let $\Sigma$ be a hypersurface of $M$, equipped with an ``outward'' normal.  We define\footnote{Our sign convention is that a sphere in Euclidean space with its usual outward pointing normal has $H>0$.} 
 the (outward) \emph{null expansion} $\theta^+_\Sigma := H_\Sigma+\tr_\Sigma k$, and we write $\theta^+$ if the context is clear. We say that $\Sigma$ is 
\begin{itemize}
\item \emph{outer trapped} if $\theta^+ < 0$,
\item \emph{weakly outer trapped} if $\theta^+ \le 0$,
\item \emph{marginally outer trapped}, or a \emph{MOTS} if $\theta^+ = 0$,
\end{itemize}
at every point of $\Sigma$. Recall that if $(M, g, k)$ sits inside a  spacetime, \textit{i.e.}\ a time-oriented Lorentzian manifold, $(N, \mathbf{g})$, then $\theta^+$ can also be thought of as the trace of the null second fundamental form of $\Sigma$ with respect to a future-directed outward null normal, and from this perspective, $\theta^+$ depends only on how $\Sigma$ embeds into $N$ and not on choice of $M$.

As there are various inequivalent definitions of asymptotic flatness, we select the following one (which implies completeness):
\begin{definition}\label{def:AF}
Let $n\ge3$. We say that an initial data set $(M^n, g, k)$ is \emph{asymptotically flat} if there exists a compact set $K \subset M$ such that $M\smallsetminus K$ is a disjoint union of ends, each of which is diffeomorphic to $\rr^n\smallsetminus B$ for some closed ball $B\subset \rr^n$, and in each of these coordinate charts, 
\begin{align*}
 g_{ij}(x)&=\delta_{ij} + O_2(|x|^{-q}) \\
 k_{ij}(x)&= O_1(|x|^{-q-1}),
\end{align*}
for some $q>(n-2)/2$, and also
 \[ (\mu, J)\in L^1(M).\]
\end{definition}
Each end of an asymptotically flat initial data set has a well-defined \emph{ADM energy-momentum} $(E, P)$, whose definition statement is not essential to the exposition in this note. Theorem 1 of~\cite{EichmairHuangLeeSchoen} states the following:

\begin{thm}\label{thm:EHLS}
Let $n<8$, and let $(M^n, g, k)$ be an asymptotically flat initial data set without boundary. If $(M, g, k)$ satisfies the dominant energy condition everywhere, then the ADM energy momentum $(E,P)$ of each end of $(M, g, k)$ satisfies $E\ge |P|$. 
\end{thm}
\begin{remark}\label{rmk:Holder}
Technically, as stated, Theorem 1 of~\cite{EichmairHuangLeeSchoen} requires some H\"{o}lder decay of $(\mu, J)$ rather than mere integrability (see Definition 3 of~\cite{EichmairHuangLeeSchoen}), but it was observed in~\cite{Lee:book} that this is unnecessary. (See footnotes on pages 255 and 299.) In fact, we will explain below exactly what (small) modification must be made to the argument.
\end{remark}

Our main result is the following:
\begin{thm}\label{thm:main}
Let $n<8$, and let $(M^n, g, k)$ be an asymptotically flat initial data set with boundary. If $(M, g, k)$ satisfies the dominant energy condition everywhere, and the boundary is outer trapped, then the ADM energy momentum $(E,P)$ of each end of $(M, g, k)$ satisfies $E\ge |P|$. 
\end{thm}
In the above statement, ``outer trapped'' means with the respect to the normal pointing ``outward'' toward the designated end, which means that the normal points \emph{into} the manifold $M$.

\begin{proof}
Without loss of generality, assume that $(M, g, k)$ has only one end: If it has more than one end, then large coordinate spheres in all of the other ends are outer trapped with respect to the end we are interested in, and by cutting off those other ends, we obtain a one-ended manifold with an outer trapped boundary (now with more components).

Next observe that there exists a smooth compact manifold whose boundary is diffeomorphic to $\partial M$. (For example, just take $M$ itself and compactify the infinite end.) Let $\hat{M}$ be the result of smoothly gluing this smooth compact manifold to $M$ along their diffeomorphic boundaries. Now smoothly extend $g$ and $k$ \emph{arbitrarily} from $M$ to all of $\hat{M}$ to obtain a new asymptotically flat initial data set $(\hat{M}, g, k)$ with one end and no boundary. Note that $(g,k)$ satisfies the DEC on the subset $M\subset \hat{M}$, and that $\partial M\subset \hat{M}$ is outer trapped, but essentially nothing is known about $(g,k)$ away from~$M$. 

Our goal is to produce a small perturbation $(\hat{M}, \tilde{g}, \tilde{k})$ with the property that $(\tilde{g}, \tilde{k})$ has \emph{harmonic asymptotics} in the sense of~\cite{EichmairHuangLeeSchoen, Lee:book} and $(\tilde{g}, \tilde{k})$ satisfies the \emph{strict} DEC on $M$. Essentially, this follows from the proof of the density theorem in~\cite{EichmairHuangLeeSchoen}. The only difference is that instead of obtaining the strict DEC everywhere (as in~\cite{EichmairHuangLeeSchoen}), we will only obtain the strict DEC at points where the original $(g,k)$ satisfied the DEC, namely, the subset $M$. We will review some of the steps of this construction below.
Since being outer trapped is an open condition (with respect to the topology used for the small perturbation), it follows that $\partial M\subset \hat{M}$ will remain outer trapped with respect to $(\tilde{g}, \tilde{k})$ as long as the perturbation from $(g, k)$ to  $(\tilde{g}, \tilde{k})$  is small enough. Moreover,  the ADM energy momentum $(\tilde{E}, \tilde{P})$ of $(\tilde{g}, \tilde{k})$ can be taken to be arbitrarily close to the original values $(E, P)$ by choosing a small enough perturbation.

From here we can follow the exact same proof of the positive mass theorem that was given in~\cite{EichmairHuangLeeSchoen} for initial data sets with strict DEC and harmonic asymptotics, for $(\hat{M}, \tilde{g}, \tilde{k})$, to see that $\tilde{E}\ge|\tilde{P}|$. The point is that the proof involves the construction of MOTS (using~\cite{Eichmair2009}), but the outer trapped boundary $\partial M$ now acts as a barrier so that all of the MOTS will lie in $M$, where the strict DEC holds. The region of $\hat{M}$ away from $M$, where we have no control over the DEC, turns out to be irrelevant. 

Here we will describe the perturbation procedure in more detail, mainly for the convenience of the reader. We will follow the exposition given in~\cite[Chapter 9]{Lee:book}, which the reader may consult for further explanation of the ideas used here. For convenience, we use the quantity $\pi_{ij} :=k_{ij}-(\tr_g k) g_{ij}$ in place of $k$. The \emph{constraint operator} on initial data is defined by 
\[ \Phi(g,\pi) := (2\mu, J^i) = \left( R_g +\tfrac{1}{n-1} (\tr_g \pi)^2 -|\pi|_g^2, (\Div_g \pi)^i \right),\]
and the \emph{modified constraint operator} at $(g,\pi)$, first introduced in~\cite{CorvinoHuang}, is defined by
\[ \overline{\Phi}_{(g,\pi)} (\gamma, \tau) = \Phi(\gamma, \tau)  + \left(0, \tfrac{1}{2} g^{ij}\gamma_{jk} J^k\right),\]
where $J^k= (\Div_g \pi)^k$.

The main usefulness of the modified constraint operator is that, unlike the regular constraint operator, it allows us to control the DEC. Specifically, for any $(\bar{g}, \bar{\pi})$, if
\[  \overline{\Phi}_{(g,\pi)} (\bar{g}, \bar{\pi})  = \overline{\Phi}_{(g,\pi)} (g, \pi) + (\psi, 0),\] 
for some function $\psi$, then as long as $|\bar{g}-g|_g\le 3$,
\begin{equation}
 |\bar{J}|^2_{\bar{g}} \le |J|_g^2,\label{J-compare}
 \end{equation}
where $\bar{J}^i= (\Div_{\bar{g}} \bar{\pi})^i$.  See~\cite[Lemma 9.15]{Lee:book} for a proof. (In contrast, there is no reason for this inequality to hold if one replaces $\overline{\Phi}_{(g,\pi)}$  by $\Phi$ above.)

We will actually perform two perturbations on the initial data $(g,\pi)$ defined on the manifold $\hat{M}$. The first perturbation is designed to obtain $(\bar{g}, \bar{\pi})$ satisfying the $(1+t)$-strict DEC on $M$ with some $t>0$, meaning that $\bar{\mu} > (1+t)|\bar{J}|_{\bar{g}}$. This is essentially~\cite[Theorem 22]{EichmairHuangLeeSchoen} or~\cite[Lemma 9.17]{Lee:book}.
Let $f$ be a positive function on $\hat{M}$ such that $f$ decays exponentially at infinity,  $f\le 1$ everywhere, and $f=1$ on some large coordinate ball containing the compact part of $\hat{M}$. For small $t>0$, we look for a solution $(\bar{g}, \bar{\pi})$ to the equation
\[  \overline{\Phi}_{(g,\pi)} (\bar{g}, \bar{\pi})  = \overline{\Phi}_{(g,\pi)} (g, \pi) + (2t(f+|J|_g), 0),\] 
such that $(\bar{g}, \bar{\pi})$ is a small perturbation of $(g,\pi)$.\footnote{Consult the references given to see the precise function spaces and the appropriate notions of closeness used in these arguments.}
 If we can do this, then 
\[\bar{\mu} = \mu + t(f+|J|_g) > \mu + t|J|_g = (\mu-|J|_g) + (1+t)|J|_g \ge (\mu-|J|_g) + (1+t)|\bar{J}|_{\bar{g}},\] 
where we used~\eqref{J-compare} for the last inequality. From this we can see that at every point where the DEC $\mu-|J|_g\ge0$ holds, we obtain the desired inequality $\bar{\mu} > (1+t)|\bar{J}|_{\bar{g}}$, so this inequality holds on all of $M$.  To prove existence of $(\bar{g}, \bar{\pi})$ for sufficiently small $t>0$, one can use the fact that $\left.D\overline{\Phi}_{(g,\pi)}\right|_{(g,\pi)}$ is surjective~\cite[Theorems 9.16]{Lee:book}.  This was proved for the regular constraint operator in~\cite{CorvinoSchoen} (see also~\cite[Theorems 9.9]{Lee:book}), and the proof for the modified constraint operator is essentially the same. 

The second perturbation is designed to impose so-called harmonic asymptotics while keeping the strict DEC on $M$. Let $\varphi$ be a cutoff function that is exactly $1$  on some large coordinate ball containing the compact part of $\hat{M}$ and vanishes outside some larger ball, and define $\varphi_k(x):=\varphi\left(\tfrac{x}{k} \right)$. Let $f$ be the same exponentially decaying function as above. For large $k$, we look for a solution
 $(\tilde{g}, \tilde{\pi})$ to the equation 
\[ \Phi(\tilde{g}, \tilde{\pi}) = \varphi_k\Phi(\bar{g}, \bar{\pi})+\left(\tfrac{2}{k}f,0\right)\]
such that   $(\tilde{g}, \tilde{\pi})$ is a small perturbation of $(\bar{g}, \bar{\pi})$ and also has harmonic asymptotics. If we can do this, then we can see that at every point of $M$, we have
\[
\tilde{\mu} = \varphi_k \bar{\mu}+ \tfrac{1}{k}f
 > \varphi_k (1+t)|\bar{J}|_{\bar{g}} 
\ge \varphi_k |\bar{J}|_{\tilde{g}}
=|\tilde{J}|_{\tilde{g}},\]
where the last inequality holds as long as $\tilde{g}$ is close enough to $\bar{g}$ 
so that the difference between $|\bar{J}|_{\bar{g}}$ and $|\bar{J}|_{\tilde{g}}$ can be absorbed by the $t|\bar{J}|_{\bar{g}}$ term. Therefore the strict DEC holds on $M$.

To solve for $(\tilde{g}, \tilde{\pi})$, surjectivity of $D\Phi|_{(\bar{g},\bar{\pi})}$ is not enough since we also want $(\tilde{g}, \tilde{\pi})$ to have harmonic asymptotics. For our purposes, the definition of harmonic asymptotics is not as important as the fact that $(\tilde{g}, \tilde{\pi})$ has harmonic asymptotics if 
we can write 
\[ (\tilde{g}, \tilde{\pi}) =\left( u^{\frac{4}{n-2}}g_{\mathbb{E}}, u^{\frac{2}{n-2}} (\mathfrak{L}_{g_{\mathbb{E}}} Y)\right),\]
outside some compact set, for some positive function $u$
 and some vector field~$Y$, and if we also know that  $\Phi(\tilde{g}, \tilde{\pi})$ decays sufficiently fast. Here, $g_{\mathbb{E}}$ is the Euclidean metric, and $\mathfrak{L}_g Y:=L_Y g - (\Div_g Y)g$.
See~\cite[Lemma 9.8]{Lee:book} for why such a solution must have harmonic asymptotics.
 Note that our $\Phi(\tilde{g}, \tilde{\pi})$ has been chosen to decay exponentially, which is more than fast enough for~\cite[Lemma 9.8]{Lee:book} to apply.

The desired solution exists by work of~\cite{CorvinoSchoen}. We summarize the argument: For large $\lambda>0$, let $(g_\lambda, \pi_\lambda)$ be initial data that interpolates between $(\bar{g},\bar{\pi})$ in the ball of radius of $\lambda$ and $(g_{\mathbb{E}}, 0)$ outside the ball of radius $2\lambda$ and consider the operator
\[ T_\lambda(u, Y) := \Phi\left( u^{\frac{4}{n-2}}g_{\lambda},u^{\frac{2}{n-2}} (\pi_\lambda+ \mathfrak{L}_{g_{\lambda}} Y )\right).\]
One can show that $DT_\lambda|_{(1,0)}$ is an elliptic operator. For large $\lambda$ and $k$, $\varphi_k\Phi(\bar{g}, \bar{\pi})+\left(\tfrac{2}{k}f,0\right)$ is very close to $T_\lambda(1,0)$ in the appropriate norm, so if $DT_\lambda|_{(1,0)}$ were surjective, we could solve $T_\lambda(u,Y)= \varphi_k\Phi(\bar{g}, \bar{\pi})+\left(\tfrac{2}{k}f,0\right)$ as desired. Although  $DT_\lambda|_{(1,0)}$  might have a finite dimensional cokernel, surjectivity of $D\Phi|_{(g_\lambda,\pi_\lambda)}$ together with the inverse function theorem implies that we can solve $T_\lambda(u,Y)=\varphi_k\Phi(\bar{g}, \bar{\pi})+\left(\tfrac{2}{k}f,0\right)$ ``modulo'' a finite dimensional space of smooth \emph{compactly supported} tensors. That is, there exists smooth, compactly supported $(h,w)$ such that
 
 \[ \Phi\left( u^{\frac{4}{n-2}}g_{\lambda}+h, u^{\frac{2}{n-2}}(\pi_\lambda+ \mathfrak{L}_{g_{\lambda}} Y)+w\right) =\varphi_k\Phi(\bar{g}, \bar{\pi})+\left(\tfrac{2}{k}f,0\right).\]
Thus $(\tilde{g}, \tilde{\pi})=\left( u^{\frac{4}{n-2}}g_{\lambda}+h, u^{\frac{2}{n-2}}(\pi_\lambda+ \mathfrak{L}_{g_{\lambda}} Y)+w\right)$ is our desired solution. See~\cite[Theorem 9.10, Proposition 9.11]{Lee:book} for a more detailed exposition.
\end{proof}

\begin{remark}\label{rmk:Holder2}
Following up on Remark~\ref{rmk:Holder}, in the proof above, we use slightly different prescribed values for $\Phi(\tilde{g}, \tilde{\pi})$ from what is used in the original proof of~\cite[Theorem 18]{EichmairHuangLeeSchoen} (see pages 118--119), and this is the reason why the H\"{o}lder decay assumption on $(\mu, J)$ in \cite{EichmairHuangLeeSchoen} is unnecessary. Basically, we need H\"{o}lder decay of $\Phi(\tilde{g}, \tilde{\pi})$ in order to apply~\cite[Lemma 9.8]{Lee:book}, but we do not need H\"{o}lder decay of $\Phi(\bar{g}, \bar{\pi})$ to make this work.
 \end{remark}

We will now discuss some consequences of Theorem~\ref{thm:main}.

\begin{cor}\label{wot1}
Let $n<8$, and let $(M^n, g, k)$ be an asymptotically flat initial data set with boundary, satisfying the dominant energy condition everywhere. If the boundary is \emph{weakly} outer trapped ($\theta^+ \leq 0$ everywhere) and each component of the boundary has a point where $\theta^+ < 0$, then the ADM energy momentum $(E,P)$ of each end of $(M, g, k)$ satisfies $E\ge |P|$. 
\end{cor} 

\begin{proof}  By \cite[Lemma 5.2]{AndMetz}, $\partial M$ can be perturbed  to a surface $\Sigma$ in $M$, which is isotopic to $\partial M$ and  outer trapped. Now apply Theorem~\ref{thm:main} to the initial data set $(M',g|_{M'}, k|_{M'})$, where $M' \subset M$ is obtained by ``truncating" $M$ at~$\Sigma$.\end{proof}

For the next result, consider the situation of an initial data set $(M, g, k)$ sitting inside a spacetime
$(N, \mathbf{g})$, and suppose $\Sigma$ is a compact
hypersurface of $M$ with an outward-pointing normal. Then we can define a future-directed outward null normal $\ell_+$  to $\Sigma$ in $N$, which generates null geodesics  
$\gamma$ emanating from $\Sigma$, and we obtain a smooth family $\Sigma_t$  by flowing along these geodesics for small affine time $t$, 
defining $\ell_+ = \gamma'$ along $\Sigma_t$. Recall that the Raychaudhuri equation states that the null expansion of $\Sigma_t$ in $N$
evolves in $t$ according to the equation 
\[  (\theta^+)' = \tfrac{-1}{n-1}(\theta^+)^2 - |\sigma|^2 - \ric(\ell_+, \ell_+),\]
where $\sigma$ is the shear tensor of $\Sigma_t$ (that is, the trace-free part of the null second fundamental form of $\Sigma_t$ in $N$) and $\ric$ is the Ricci curvature of $\mathbf{g}$.

\begin{cor}\label{wot2} 
Let $n<8$, and let $(M^n, g, k)$ be an asymptotically flat initial data set with weakly outer trapped boundary, sitting inside a  spacetime $(N, \mathbf{g})$ which satisfies (the spacetime version of) the dominant energy condition. Suppose also that on each component of $\partial M$,
either the shear tensor $\sigma$ or the curvature quantity $\ric(\ell_+,\ell_+)$ is not identically zero. Then the ADM energy momentum $(E,P)$ of each end of $(M, g, k)$ satisfies $E\ge |P|$. 
\end{cor}  

\begin{proof}  Take $\Sigma=\partial M$ and consider the family $\Sigma_t$ in $N$ described above. The dominant energy condition on $\mathbf{g}$ implies that $ \ric(\ell_+, \ell_+) \ge 0$, so the Raychaudhuri equation tells us that for all $t$, $(\theta^+)'(t) \le0$ everywhere, and our hypotheses imply that $(\theta^+)'(0) <0$ somewhere on each component of $\Sigma$. So for small $t$, $\Sigma_t$ has $\theta^+\le0$ and $\theta^+<0$ somewhere on each component of $\Sigma_t$. 

Next we slightly deform the initial data set $(M,g,k)$  in a spacetime neighborhood of $\partial M$ in $N$ to produce an initial data set $(M',g',k')$ sitting inside $N$ with $\partial M' = \Sigma_t$. Since $(N, \mathbf{g})$ satisfies the (spacetime) dominant energy condition, $(M',g',k')$ satisfies the DEC for initial data sets, and we can now apply Corollary \ref{wot1} to $(M',g',k')$, which has the same ADM energy-momentum as $(M, g, k)$. 
\end{proof}

The main improvement of Corollary \ref{wot2} over Corollary \ref{wot1} is that it applies to the case when the boundary is a MOTS, but the tradeoff is that it requires an ambient spacetime. The next corollary is a pure initial data result that applies to the case of MOTS boundary but only under certain additional conditions.

Recall that MOTS admit an important notion of stability (see \cite{AndMarsSim2},  \cite[Section 7.5]{Lee:book}). Given a MOTS $\Sigma$ in $(M,g,k)$, let 
$L_{\S}$ be the {\it MOTS stability operator}, which is essentially the linearization of $\theta^+$ with respect to normal first variations of $\Sigma$. For a compact MOTS $\Sigma$, there exists an eigenvalue of $L_{\S}$ with minimal real part, which is real and denoted by $\l_1(L_{\S})$, and we 
say that $\S$ is \emph{stable} if $\l_1(L_{\S}) \ge 0$, and \emph{strictly stable} if $\l_1(L_{\S}) >  0$. If $\Sigma$ is connected, there exists a principal eigenfunction, which is a positive eigenfunction with eigenvalue  $\l_1(L_{\S})$.
Recall that for $n<8$, if an asymptotically flat initial data set has a weakly outer trapped boundary, then for each end, there exists a MOTS which is \emph{outermost} with respect to that end, and that outermost MOTS must be stable (see \cite{AnderssonEichmairMetzger} and \cite[Theorem 7.40]{Lee:book}). In other words, if we want to relax the \emph{outer trapped}  assumption of Theorem~\ref{thm:main} to \emph{weakly outer trapped}, it is sufficient to consider the case of stable MOTS boundary. The next corollary is the closest we get to this.

\begin{cor}\label{strictlystable}
Let $n<8$, and let $(M^n, g, k)$ be an asymptotically flat initial data set with boundary. Assume that $(M, g, k)$ satisfies the dominant energy condition everywhere, and also that $M$ contains a strictly stable MOTS in its interior which is isotopic to $\partial M$. Then the ADM energy momentum $(E,P)$ of that end satisfies $E\ge |P|$. 
\end{cor} 

We note that a similar assumption is used in the proof of the main result in \cite{AndDahlGalPollack}.  

In \cite{Mars}, M. Mars  obtained interesting criteria for a MOTS to be strictly stable.  Let $\S$ be a compact  ``cross section'' of an event horizon $\calH$ in a spacetime $(N, \mathbf{g})$.  Suppose that $\calH$ is a {\it Killing horizon} (as discussed in \cite{Chru:bhbook}, for example) or, more generally, what Mars calls a ``non-evolving'' horizon.  Since such horizons are totally geodesic, it follows that cross sections are MOTS.  Mars shows, under further mild conditions, that if the {\it surface gravity} of $\calH$ is positive, then $\S$ is a strictly stable MOTS. (See \cite[Proposition 3]{Mars} for a more detailed statement.) Therefore Corollary \ref{strictlystable} applies to initial data sets with boundary that may arise from this natural class of spacetimes studied by Mars.

\begin{proof}[Proof of Corollary \ref{strictlystable}]  Let $\Sigma$ be a strictly stable MOTS enclosing $\partial M$, and for now assume that $\Sigma$ is connected. Let $\varphi$ be the principal eigenfunction of $L_{\S}$, meaning that $\varphi>0$ and  $L_{\S}(\varphi) = \l_1(L_{\S})  \varphi$. Choose any variation  $\S_t$ of $\S = \S_0$ with variation vector field $X = -\varphi \nu$ at $\S$, where $\nu$ is the outward unit normal to $\S$. 
By the definition of $L_{\S}$, the null expansion $\theta^+$ of $\Sigma_t$ satisfies
\[
\left. \frac{\partial \th^+}{\partial t} \right |_{t=0}   = L_{\S}(-\varphi) = -  \l_1(L_{\S})  \varphi < 0,
\]
since $\S$ is strictly stable. So for small $t>0$, $\Sigma_t$ is outer trapped and is isotopic to $\partial M$. If $\Sigma$ is not connected, then each component is also strictly stable, and we can just run the same argument on each component to obtain the same result.
Now apply Theorem~\ref{thm:main} to the initial data set $(M',g|_{M'}, k|_{M'})$, where $M' \subset M$ is obtained by ``truncating" $M$ at $\S_t$.
\end{proof}

We can also obtain a corollary that applies to general MOTS boundaries if we assume a ``no KIDs'' condition. Recall that a \emph{KID} on an initial data set $(\Omega, g, k)$ is a nontrivial element in the kernel of $\left.D\Phi\right|_{(g,\pi)}^*$, which is the formal adjoint of the linearization of the constraint map $\Phi$ described in the proof of Theorem~\ref{thm:main}.  For a given set $S$ in an initial data set $(M, g, k)$, we will say that $(M, g, k)$ \emph{has no local KIDs near} $S$ if the kernel of $\left.D\Phi\right|_{(g,\pi)}^*$ over $\Omega$ is trivial for every open set $\Omega$ containing $S$.
Because this kernel condition is heavily overdetermined, the existence of a KID is a very special situation, and in fact, it was shown that having no local KIDs at all is a generic condition within the space of vacuum initial data sets~\cite{BeigChruscielSchoen}.

\begin{cor}\label{cor:KIDs}
Let $n<8$, and let $(M^n, g, k)$ be an asymptotically flat initial data set with boundary, satisfying the dominant energy condition everywhere. Assume that the boundary is a MOTS, that there are no local KIDs near the boundary, and that we either have $J=0$ or $\mu>|J|_g$ near the boundary. 
Then the ADM energy momentum $(E,P)$ of each end of $(M, g, k)$ satisfies $E\ge |P|$. 
\end{cor}
\begin{proof}
Under the exact hypotheses of the corollary, the main result of~\cite{ChruscielGalloway} says that there exists a perturbation $(\tilde{g}, \tilde{k})$ of $(g,k)$ such that $(\tilde{g}, \tilde{k})$ satisfies the DEC everywhere and is identically equal to $(g,k)$ away from some neighborhood of $\partial M$, and moreover 
$\partial M$ is outer trapped in $(M, \tilde{g}, \tilde{k})$. The result now follows directly from Theorem~\ref{thm:main}.
\end{proof}
\begin{remark}
The condition that $J=0$ or $\mu>|J|_g$ near the boundary (which comes from~\cite{ChruscielGalloway}) is a consequence of the fact that the kernel of $\left.D\Phi\right|_{(g,\pi)}^*$ is related to the prescribed constraint equation, and as noted earlier, prescribing the constraints does not allow one to control the DEC in general. We conjecture that if one were to replace the ``no KIDs'' hypothesis by an analogous ``modified'' hypothesis related to $\left.D\overline{\Phi}_{(g,\pi)}\right|_{(g,\pi)}^*$, then one can use the results of~\cite{CorvinoHuang} in the argument in~\cite{ChruscielGalloway} to remove the condition that $J=0$ or $\mu>|J|_g$ near the boundary. In fact, invoking the results of~\cite{HuangLee} could narrow down the only possible ``exceptions'' to the positive mass theorem with MOTS boundary to cases where a neighborhood of the boundary sits inside a null perfect fluid spacetime with constant pressure.
\end{remark}

\begin{remark}
Recall that an \emph{inner untrapped} (\textit{i.e.}\ \emph{``past'' outer trapped}) hypersurface $\Sigma$ in an initial data set is one with $\theta^-_\Sigma := -H_\Sigma + \tr_\Sigma k>0$. Simply by flipping the sign of $k$, one trivially sees that all of the results of this paper still hold if one replaces all instances of ``(weakly) outer trapped'' by ``(weakly) inner untrapped'' and all ``MOTS'' by ``MITS'' (\textit{i.e.}\ hypersurfaces with $\theta^-=0$), as well as other corresponding changes.
 It is slightly less obvious, but still true and easy to check, that in all of the results above, one can relax the assumptions so that each \emph{component} of the relevant hypersurface is ``either (weakly) outer trapped or (weakly) inner untrapped,'' or is ``either a MOTS or a MITS,'' depending on context, together with other obvious changes.
\end{remark}

\smallskip
\noindent
\textsc{Declaration of Funding.}  
The research of GJG was supported by the NSF under the grant DMS-171080.

\bibliographystyle{amsplain}
\bibliography{PMT}

\end{document}